\documentclass[proof]{WileyASNA-v1}

\usepackage{graphicx}

\usepackage{color}
\usepackage{xspace}

\newcommand{\ktwo}{{\it K2}\xspace}
\newcommand{\kep}{\textit{Kepler}\xspace}
\newcommand{\hd}{HD\,106315\xspace}
\newcommand{\feh}{\mbox{[Fe/H]}\xspace}
\newcommand{\teff}{\mbox{$T_{\rm *, eff}$}\xspace}
\newcommand{\logg}{\mbox{$\log g_*$}\xspace}

\articletype{Original Article}%

\received{31 January, 2020}
\revised{13 February, 2020}
\accepted{<day> <Month>, <year>}

\begin{document}

\title{Shallow transit follow-up from NGTS: simultaneous observations of HD106315 with 11 identical telescopes}

\author[1]{Alexis M. S. Smith}
\author[1]{Philipp Eigm\"uller}
\author[1,2]{Ramanathan Gurumoorthy}
\author[1]{Szil\'ard Csizmadia}
\author[3,4]{Daniel Bayliss}
\author[5]{Matthew R. Burleigh}
\author[1]{Juan Cabrera}
\author[5]{Sarah L. Casewell}
\author[1]{Anders Erikson}
\author[5]{Michael R. Goad}
\author[5]{Andrew Grange}
\author[6,7]{James S. Jenkins}
\author[3,4]{Don Pollacco}
\author[1,8,9]{Heike Rauer}
\author[5]{Liam Raynard}
\author[10]{St\'ephane Udry}
\author[3,4]{Richard G. West}
\author[3,4]{Peter J. Wheatley}

\authormark{SMITH \textsc{et al}}

\address[1]{Institut f\"ur Planetenforschung, Deutsches Zentrum f\"ur Luft- und Raumfahrt (DLR), Berlin, Germany}

\address[2]{Faculty of Aerospace Engineering, Delft University of Technology, Delft, The Netherlands}
\address[3]{Department of Physics, University of Warwick, Coventry, UK}
\address[4]{Centre for Exoplanets and Habitability, University of Warwick, Coventry, UK}
\address[5]{School of Physics and Astronomy, University of Leicester, Leicester, UK}
\address[6]{Departamento de Astronom\'ia, Universidad de Chile, Santiago, Chile}
\address[7]{Centro de Astrof\'isica y Tecnolog\'ias Afines (CATA), Santiago, Chile}
\address[8]{Zentrum f\"ur Astronomie und Astrophysik, Technische Universit\"at Berlin, Berlin, Germany}
\address[9]{Institut f\"ur Geologische Wissenschaften, Freie Universit\"at Berlin, Berlin, Germany}
\address[10]{Department of Astronomy, University of Geneva, Versoix, Switzerland}

\corres{*Alexis M. S. Smith, Institut f\"ur Planetenforschung, Deutsches Zentrum f\"ur Luft- und Raumfahrt (DLR), Rutherfordstra\ss e 2, 12489 Berlin, Germany \email{alexis.smith@dlr.de}}


\abstract[Abstract]{The Next Generation Transit Survey (NGTS) is a photometric survey for transiting exoplanets, consisting of twelve identical 0.2-m telescopes. We report a measurement of the transit of \hd~c using a novel observing mode in which multiple NGTS telescopes observed the same target with the aim of increasing the signal-to-noise. Combining the data allows the robust detection of the transit, which has a depth less than 0.1 per cent, rivalling the performance of much larger telescopes. We demonstrate the capability of NGTS to contribute to the follow-up of \ktwo and TESS discoveries using this observing mode. In particular, NGTS is well-suited to the measurement of shallow transits of bright targets. This is particularly important to improve orbital ephemerides of relatively long-period planets, where only a small number of transits are observed from space.}



\maketitle


\section{Introduction}
\label{sec:intro}
The Next Generation Transit Survey (NGTS; \citealt{NGTS}) is the highest precision wide-field, ground-based transit survey in operation, allowing it to detect much shallower transits than the previous generation of such surveys, such as WASP \citep{Pollacco06} and HAT-Net \citep{bakos02}. To date, NGTS has discovered a number of transiting exoplanets \citep{NGTS-1,NGTS-2,NGTS-3,NGTS-5,NGTS-6,NGTS-7,NGTS-8-9,NGTS-10}, as well as probing other astrophysical phenomena, such as stellar flares \citep{Jackman_dL_flare} and low-mass eclipsing binary systems \citep{Casewell_EBLM}.

The performance of NGTS was recently demonstrated in the detection of NGTS-4b, whose $0.13 \pm 0.02$ per cent deep transit makes it the system with the shallowest transit ever discovered from the ground \citep{NGTS-4}. The detection of even shallower transits, and thus smaller planets, from space has now become routine. \ktwo, the second incarnation of NASA's \kep spacecraft \citep{K2} has discovered many such systems. In 2018, TESS (Transiting Exoplanet Survey Satellite; \citealt{TESS}) began its two-year survey of 85 per cent of the sky.

One limitation of \ktwo and TESS, however, is their observing baseline, which is typically around 80~d in the case of \ktwo, and as short as 27~d for TESS. This places strong upper limits on the orbital periods of the systems discovered by these instruments. Less than 10 per cent of the 389 planets discovered to date by \ktwo orbit with periods longer than 25 days\footnote{Statistics from NASA Exoplanet Archive \citep{NASA_exoplanet_archive}, retrieved 2019 September 19}.

\hd, also known as K2-109 ($\alpha_{\mathrm{J2000}} = 12h13m53.40s$, $\delta_{\mathrm{J2000}} = -00^{\circ}~23^\prime~36^{\prime\prime}.55$) is a system of two planets detected in Campaign 10 of \ktwo orbiting a bright (V=8.9) F5V star \citep{Rodriguez17,Crossfield17}. During observations of Campaign 10, only two transits of the outer planet `c' were observed. This not only limited the precision to which key system parameters could be determined, but resulted in a rather poorly-constrained orbital ephemeris. According to the ephemeris of \cite{Rodriguez17}, the 1~$\sigma$ uncertainty in the transit time would reach 5~h just 5.5 years after the discovery epoch. Photometric transit observations from the ground were required in order to prevent the ephemeris from being `lost' altogether, and thus impede future follow-up efforts. With this in mind, we scheduled NGTS observations of the system (Section~\ref{sec:obs}).

In addition to our NGTS observations, \cite{Lendl17} observed two transits of \hd~c from the ground with the 1.2-m Euler telescope, allowing the ephemeris to be refined and reducing the uncertainty on the orbital period by a factor of four.  Similarly, \cite{Barros17} observed one transit with one of the 1-m telescopes of the Las Cumbres Observatory. They also measured the masses of both planets orbiting \hd with 93 radial velocities from HARPS.

Observations of such a shallow transit would also prove a good test of the capabilities of NGTS in non-survey mode. In the normal mode of operation, each of the twelve NGTS telescopes observes a separate field, in order to cover the largest possible area of sky, and maximise the number of new planetary systems detected. For these observations, however, we decided to test observing the same target with all the telescopes. This is a mode of operation that we expect to prove invaluable in confirming and better-characterising shallow transits detected by NGTS itself, and in following-up shallow transits detected by TESS.

The observations of \hd offered a good opportunity to test this observing mode, and to quantify the advantages of combining data from multiple identical telescopes. This has particular relevance for the upcoming PLATO (PLAnets, Transits and Oscillations; \citealt{PLATO}) mission, which will use a total of 26 0.12-m space-based telescopes with overlapping fields-of-view to monitor a large area of sky for nearby transiting exoplanets.

The remainder of this paper is laid out as follows: in Section~\ref{sec:obs} we present our NGTS observations of \hd. In Section~\ref{sec:phot} we describe our data analysis and custom-built pipeline to produce light curves of \hd. In Section~\ref{sec:combine} we investigate the combining of data from multiple telescopes to produce a single light curve. Our discussion and conclusions can be found in Section~\ref{sec:discuss}.

\section{Observations}
\label{sec:obs}

We observed a field centered on \hd with eleven of the twelve NGTS telescopes on the night of 2017 March 08/09. Each telescope has an aperture diameter of 0.2~m. The cameras associated with these telescopes are identified within the NGTS project, and in the rest of this paper as 01, 02, 03, 06, 07, 08, 09, 10, 11, 12, and 13. In contrast to the usual NGTS survey mode, we defocussed each of the telescopes slightly in order to avoid saturation or non-linearity of the CCD response, since \hd is slightly brighter than the usual NGTS bright limit.

We note that our observations were taken on the same night as the first transit observed by \cite{Lendl17}, which was also observed by \cite{Barros17}. These observations were conducted at La Silla and Cerro Tololo, which lie to the south of the NGTS site at Paranal, by around 500~km and 600~km, respectively.

Our observations of \hd comprise around 2730 images per telescope -- more than 27\,000 in total, spanning 7.56~h. The normal survey mode for NGTS uses 10\,s exposures, but for these observations of \hd we used 7\,s exposures to further reduce the likelihood of saturation. With the fast readout time of the NGTS CCDs, this results in an observing cadence of 10\,s.

\section{Light curve generation}
\label{sec:phot}

Since our photometry is defocussed, the observations could not be reduced using the standard NGTS photometry pipeline \citep{NGTS}, used for the processing of survey observations. Instead, we developed a standalone pipeline for the processing of such datasets, based on standard aperture photometry with {\sc photutils} \citep{photutils0.3}, part of the {\sc astropy} python package \citep{astropy1}.

The major processing steps are described briefly below, and are performed on a per-camera basis. Image calibration is performed via standard bias and flat corrections as per the usual NGTS data reduction pipeline \citep{NGTS}. A master frame is generated, and {\sc SExtractor} used to perform astrometry, enabling a source catalogue to be generated, and cross-matched with UCAC4 \citep{UCAC4paper}. Aperture photometry is then performed on each source (with an aperture radius of 3.0 pixels = 15$^{\prime\prime}$, optimised to minimise the out-of-transit rms), along with background estimation via sky annuli (with inner and outer radii of 9.0 and 14.0 pixels, respectively).

The raw light curves are detrended by fitting polynomials to the airmass, and to the CCD $x$ and $y$ positions. The \hd light curve is further corrected by means of a combined reference star, consisting of the flux from five nearby stars of similar magnitude (Table~\ref{tab:comparisons}). We found the light curves of \hd generated in this way to contain little correlated noise (see Appendix~\ref{appendix}).

\begin{table}
\caption{List of comparison stars used}
\begin{center}
\begin{tabular}{lllcc} 
\toprule
ID & UCAC4 ID & $r$ mag & $J-K$ \\
\midrule
Target & 449-052646 & 9.396 & 0.263 \\
\midrule
Ref1 & 447-053330 & 8.707 & 0.675 \\
Ref2 & 449-052685 & 9.377 & 0.879 \\
Ref3 & 444-054438 & 8.663 & 0.209 \\
Ref4 & 446-054654 & 9.438 & 0.659 \\
Ref5 & 444-054448 & 10.585 & 1.03 \\
\bottomrule
\end{tabular}
\end{center}
\label{tab:comparisons}
\end{table}

~\\

\section{Measuring the planetary radius}
\label{sec:radius}
\subsection{Single-telescope light curves}
\label{sec:analysis_single}
\begin{figure*}
    \centering
    \includegraphics[width=0.8\hsize]{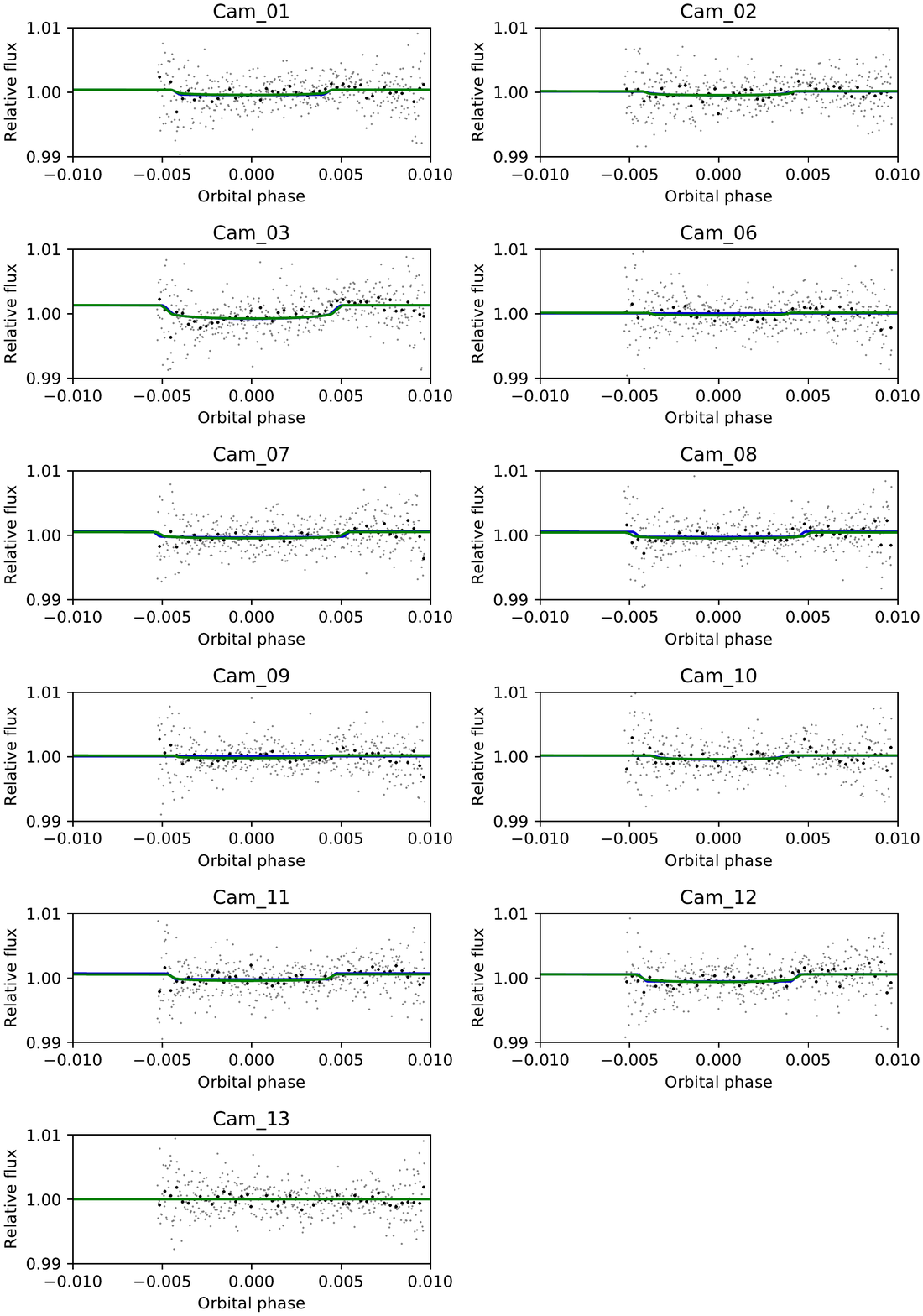}
    \caption{Light curves from individual NGTS telescopes, binned to 1~min (small grey circles) and 10~min (large black circles). The best-fitting models from a fit where $b$, $u_+$ and $u_-$ were free parameters is shown with a blue line, and from a fit where $u_+$ and $u_-$ were fixed, and $b$ was constrained is shown with a green line.
    }
    \label{fig:lcs}
\end{figure*}

We started our analysis with the eleven light curves, each the output of a different telescope/camera, the generation of which is described in Section~\ref{sec:phot}. We first fit a transit model to each light curve individually. The fits were performed with the Transit Light Curve Modeller (TLCM\footnote{http://www.transits.hu}; Csizmadia, in press), which uses MCMC for error estimation. In our first 
set of fits, the following parameters were freely fit: the planet-to-star radius ratio, $R_p/R_*$, the impact parameter, $b$, the limb-darkening coefficients, $u_+ = u_1 + u_2$ and $u_- = u_1 - u_2$, and an offset to account for possible imperfect light curve normalisation. The scaled orbital major semi-axis, $a/R_*$, was allowed to vary within the 1$\sigma$ uncertainties determined by \cite{Rodriguez17} ($a/R_* = 25.69 \pm 1.2$). The ephemeris was fixed to that of \cite{Barros17} ($P = 21.05704$~d, $t_0 = 2457569.0173$ [$\mathrm{BJD_{TDB}}$]), and the orbital eccentricity, $e$ was fixed at zero. Each MCMC run used 20 independent chains, and we used the Gelman-Rubin statistic \citep{Gelman-Rubin} to check for convergence. As a final check, the fits were repeated to check the consistency of the results, which were near identical (variations in the best-fitting parameter values were much smaller than the associated 1 $\sigma$ errors). The light curves are shown along with the fits (blue lines) in Fig.~\ref{fig:lcs}.

\begin{figure}
    \centering
    \includegraphics[width=\hsize]{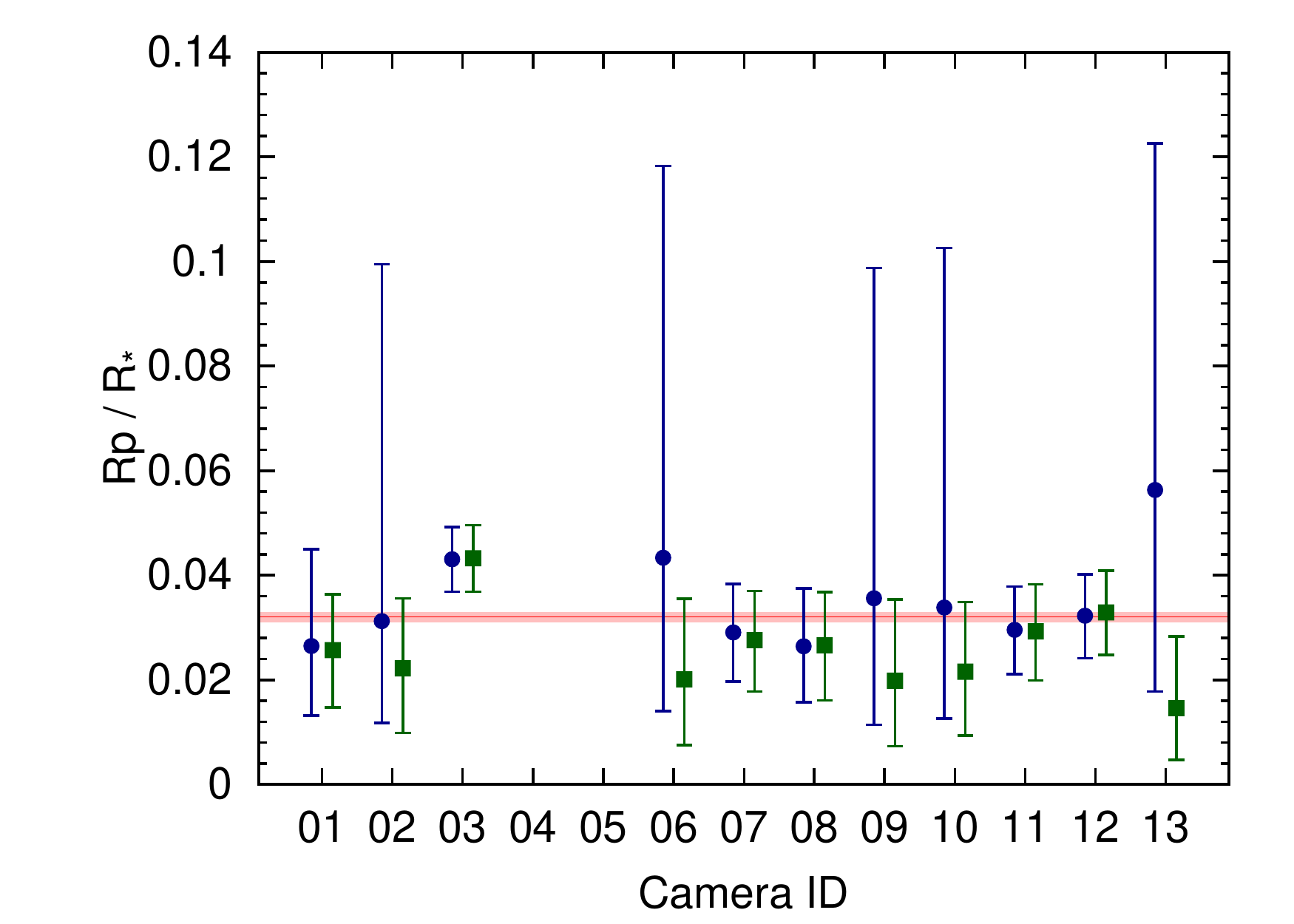}
    \caption{Fitted radius ratio for single-telescope light curves. The value and uncertainty of $R_p/R_*$ determined by \cite{Rodriguez17} is indicated with a red line. Results from the fits where $b$, $u_+$ and $u_-$ were free parameters are displayed with blue circles. Results from fits where $u_+$ and $u_-$ were fixed, and $b$ was constrained, are displayed with green squares.
    }
    \label{fig:radii}
\end{figure}

Looking at the resulting $R_p/R_*$ values (blue points, Fig.~\ref{fig:radii}), we see that in three cases (cameras 06, 09, 13) the best-fitting model is a straight line that doesn't include a transit. In these cases, the best-fitting impact parameter is larger than $R_* + R_p$, hence there is no transit. In these three cases, and for camera 10, the radius ratio is poorly constrained. In the six remaining cases, the radius ratio is reasonably well determined, and in good agreement with the value determined by \cite{Rodriguez17} (the discrepancies are $< 1 \sigma$ in all cases, except camera 03, where the discrepancy is $<2 \sigma$).

To simply things further, and to `force' the fitted model to include a transit, we decided to constrain the impact parameter to lie between 0.6 and 0.8, encompassing the best-fitting $0.688^{+0.044}_{-0.094}$ of \cite{Rodriguez17}. We also opted to fix the limb-darkening coefficients, using values from \cite{Sing09} for a star with \teff = 6250~K, \feh = -0.3, and \logg [cgs] = 4.5. The mean rms of the residuals to a single telescope fit is 2700~ppm per minute or 500~ppm per half hour.

\begin{figure}
    \centering
    \includegraphics[width=\hsize]{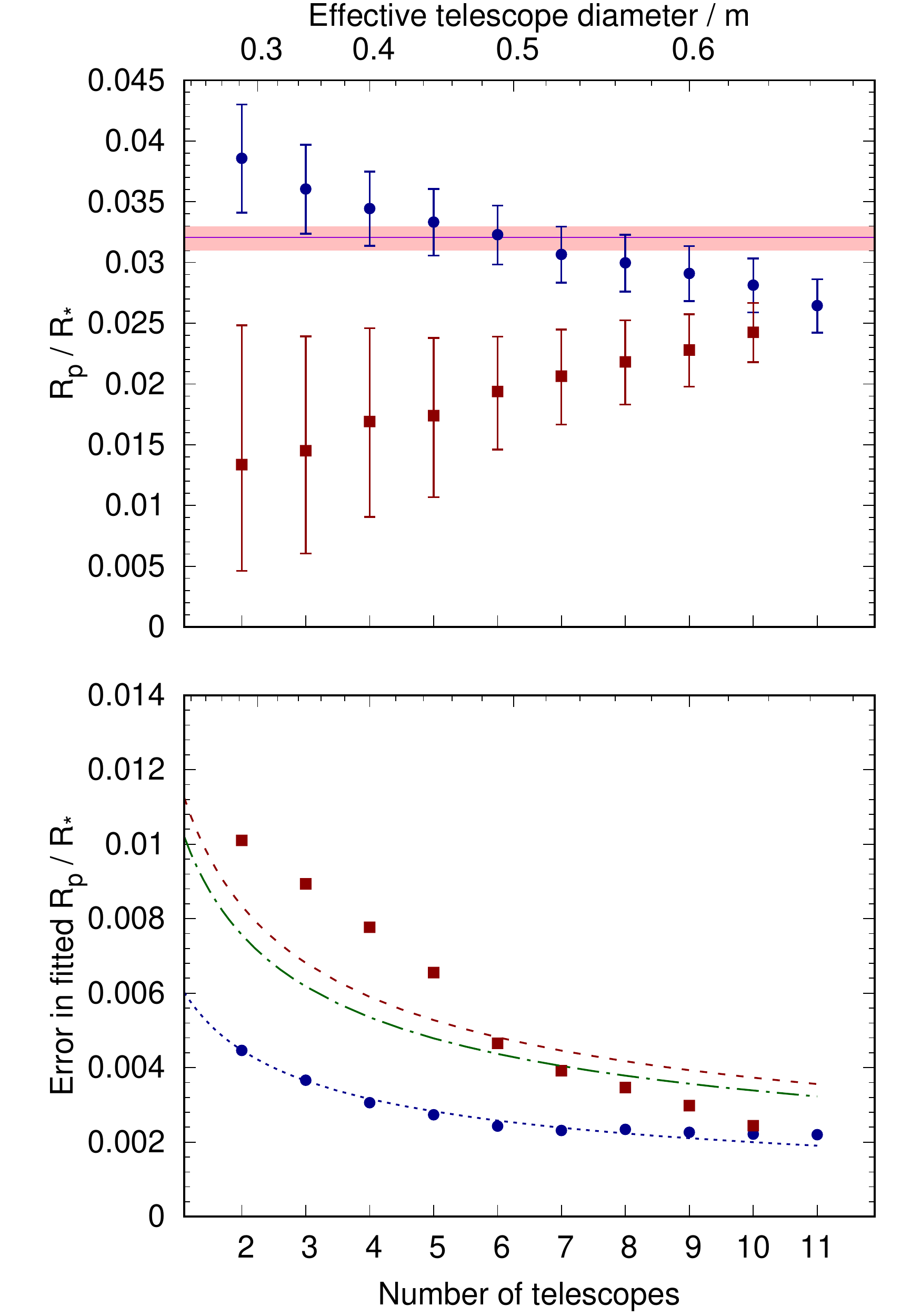}
    \caption{Upper panel: fitted radius ratio as a function of number of telescopes combined. The blue circles represent the combinations starting with the two `best' (see text) light curves and adding successively worse light curves. The red squares represent the combinations starting with the two `worst' light curves, and adding successively better light curves. The red line and shaded region indicate the best-fitting solution and 1 $\sigma$ uncertainties of \cite{Rodriguez17}. Lower panel: uncertainty on fitted radius ratio. The red squares and blue circles correspond to the same fits as the upper panel. The blue dashed line represents the expected white noise case based on the best individual light curve, the red dotted line the same, but based on the worst individual light curve, and the green dash-dotted line is based on the mean uncertainty on the radius ratio from the individual light curves.
    }
    \label{fig:combo}
\end{figure}

\subsection{Combining data from multiple NGTS telescopes}
\label{sec:combine}

After fitting the individual light curves, we resolved to fit the light curves from multiple NGTS telescopes together, to see how our determination of $R_p/R_*$ changes as the number of light curves used in the fit increases. We decided to perform two experiments, one where the light curves are added from `best-to-worst', and one in which the light curves are added from `worst-to-best'. Our ranking of the individual light curves is carried out on the basis of the magnitude of the uncertainty on $R_p/R_*$ from the individual fits with limb-darkening and impact parameter freely fitted (the blue circles in Fig~\ref{fig:radii}). Thus, camera 03 is regarded as the `best', and camera 13 as the `worst' individual light curve.

To allow for imperfect flux normalisation, we fit for a constant offset in flux between each additional light curve. Fig.~\ref{fig:combo} shows the fitted value of $R_p/R_*$ and its uncertainty, $\sigma_{R_p/R_*}$ as a function of the number of NGTS light curves included in the fit. This figure shows a gradual convergence to $R_p/R_* = 0.0264 \pm 0.0022$ when light curves from all eleven telescopes are included in the analysis. 

Furthermore, the uncertainty in the radius ratio decreases as a function of the number of telescopes, $n_\mathrm{tel}$, for both the `best-to-worst' and `worst-to-best' cases. By taking $\sigma_{R_p/R_*}$ for a single light curve, and scaling this value by $1/\sqrt{n_\mathrm{tel}}$, a comparison to the expected white noise behaviour may be made. The theoretical curves in the lower panel of Fig.~\ref{fig:combo} show such a relation for three different values of $\sigma_{R_p/R_*}$, corresponding to the smallest, largest and mean values of $\sigma_{R_p/R_*}$ from the fits to individual light curves (with limb-darkening and the impact parameter constrained; green squares in Fig.~\ref{fig:radii}). When the `best' light curve is fitted first, the data follow the white noise curve very closely, with only a slight deviation as $n_\mathrm{tel}$ approaches eleven, and the newly-included data is increasingly poor. In contrast, when we begin fitting the `worst' light curve first, the improvement in $\sigma_{R_p/R_*}$ is slow initially, but then undergoes a more rapid reduction as better data is added.

\begin{figure} 
\includegraphics[width=\hsize]{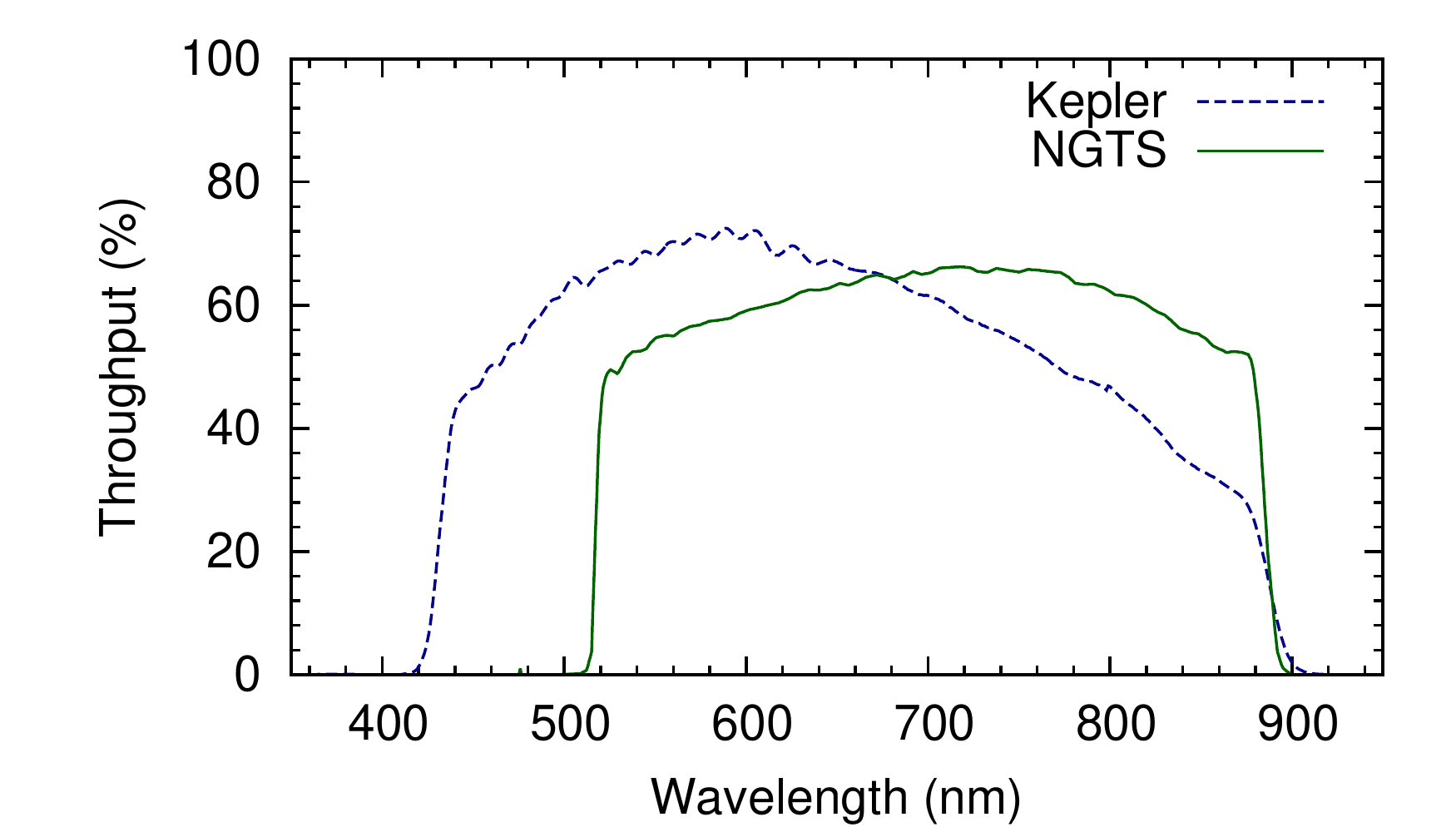} 
\caption{Comparison of \kep/\ktwo and NGTS instrument response as a function of wavelength. The Kepler data are taken from the Kepler Science Center,
and the NGTS data from \cite{NGTS}.
}
\label{fig:bands} 
\end{figure}

We note that the value of $R_p/R_*$ that our fits converge upon is somewhat smaller than that of \cite{Rodriguez17} (indicated with a red line in Fig.~\ref{fig:radii}), at a significance of 2.1~$\sigma$. One possible reason for this apparent discrepancy is the limb-darkening coefficients chosen for our fit to the NGTS data. Although the NGTS and {\it Kepler} passbands are similar (Fig.~\ref{fig:bands}), the blue cut-offs do differ markedly, and limb-darkening is stronger at these shorter wavelengths. To test this hypothesis, we tried several different approaches to choosing the limb-darkening coefficients, including a fit that allowed them to vary significantly. We observed no discernable dependence of the fitted $R_p/R_*$ on the chosen limb-darkening coefficients.

\begin{figure} 
\includegraphics[width=\hsize]{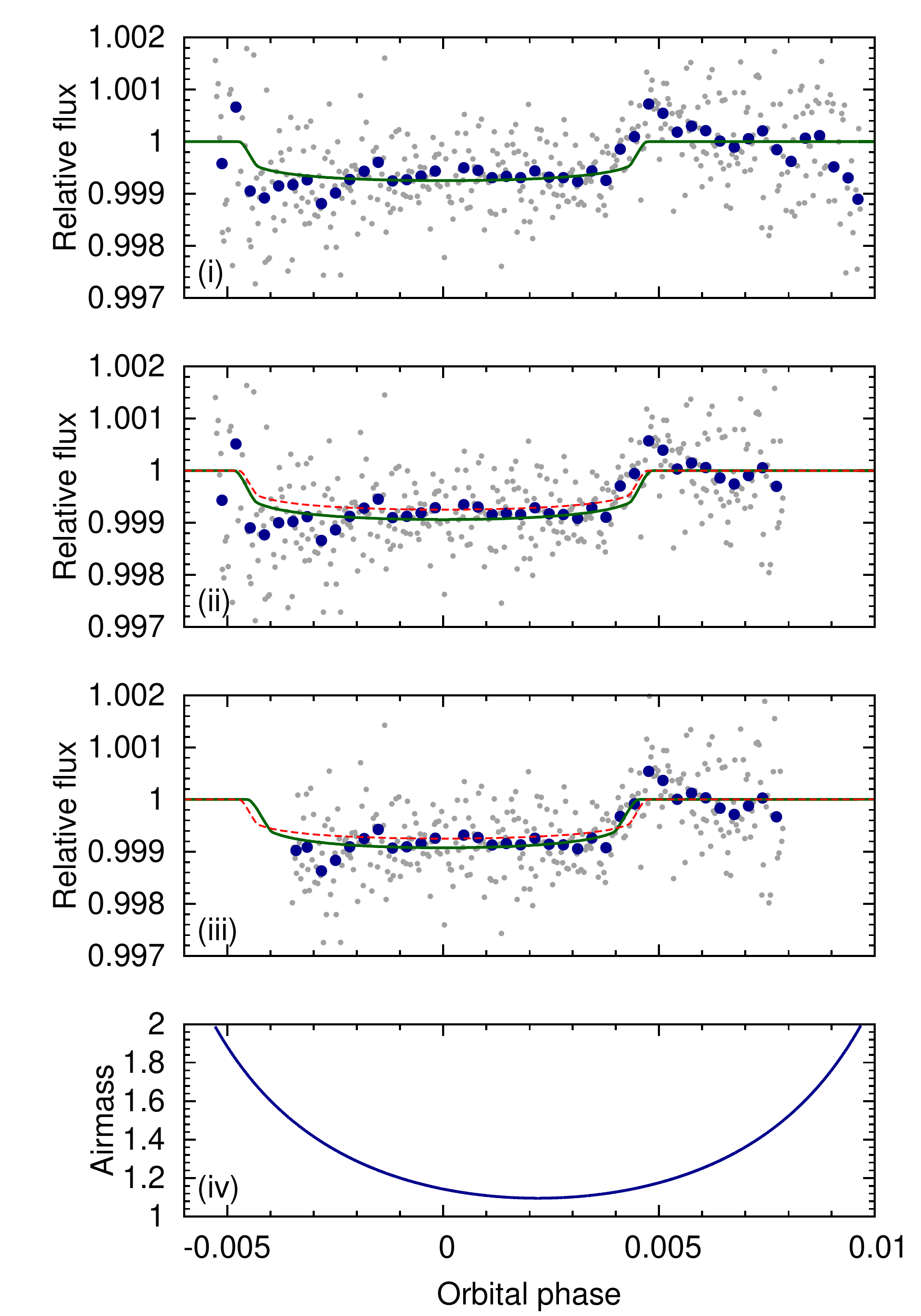} 
\caption{Combined (11 telescope) light curve of \hd. From top to bottom: (i) full light curve binned to one minute (small grey points) and ten minutes (large blue points), with best-fitting model (described in Section~\ref{sec:combine}. (ii) As top panel, but with observations made at airmass $>1.5$ excluded from the fit, which is shown with a solid green curve. The red dashed curve shows the fit from the top panel. (iii) As (ii), but with high-airmass data at the start of the night excluded as well. (iv) Target airmass during the observations.
}
\label{fig:lc_airmass} 
\end{figure}

A second potential explanation for the lower-than-expected value of $R_p/R_*$, is the difficulty in determining the out-of-transit baseline flux. In the uppermost panel of Fig.~\ref{fig:lc_airmass} we plot the combined light curve from all eleven NGTS cameras, while the bottom panel shows the airmass of \hd during the course of the night. The observations begin and end at airmass 2, meaning that all of the out-of-transit data is taken at relatively high airmass. There is almost no pre-transit data, and only a limited amount of post-transit data, significantly less than in-transit data. The light curve shows an apparent decrease in flux at the end of the night, as well as significantly increased scatter evident in the un-binned light curve, corresponding to data taken when the target was at an airmass greater than about 1.5. To test this theory, we tried fitting the light curves from all eleven telescopes, but excluding data taken at the end of the night at airmass values greater than 1.5. We also did the same, but excluding the high-airmass data from both the beginning and end of the night. The resulting light curves and best-fitting models are shown in Fig.~\ref{fig:lc_airmass}.

By excluding the high-airmass data from the end of the night, we recover a transit depth and hence $R_p/R_*$ in better agreement with the previously-published values, based on the higher-precision \ktwo light curve. The removal of additional high-airmass data, from the beginning of the night results in a virtually identical determination of $R_p/R_*$, but a slightly shorter duration transit. This results from the complete lack of data covering transit ingress in this case (Fig.~\ref{fig:lc_airmass}). 

\begin{figure} 
\includegraphics[width=\hsize]{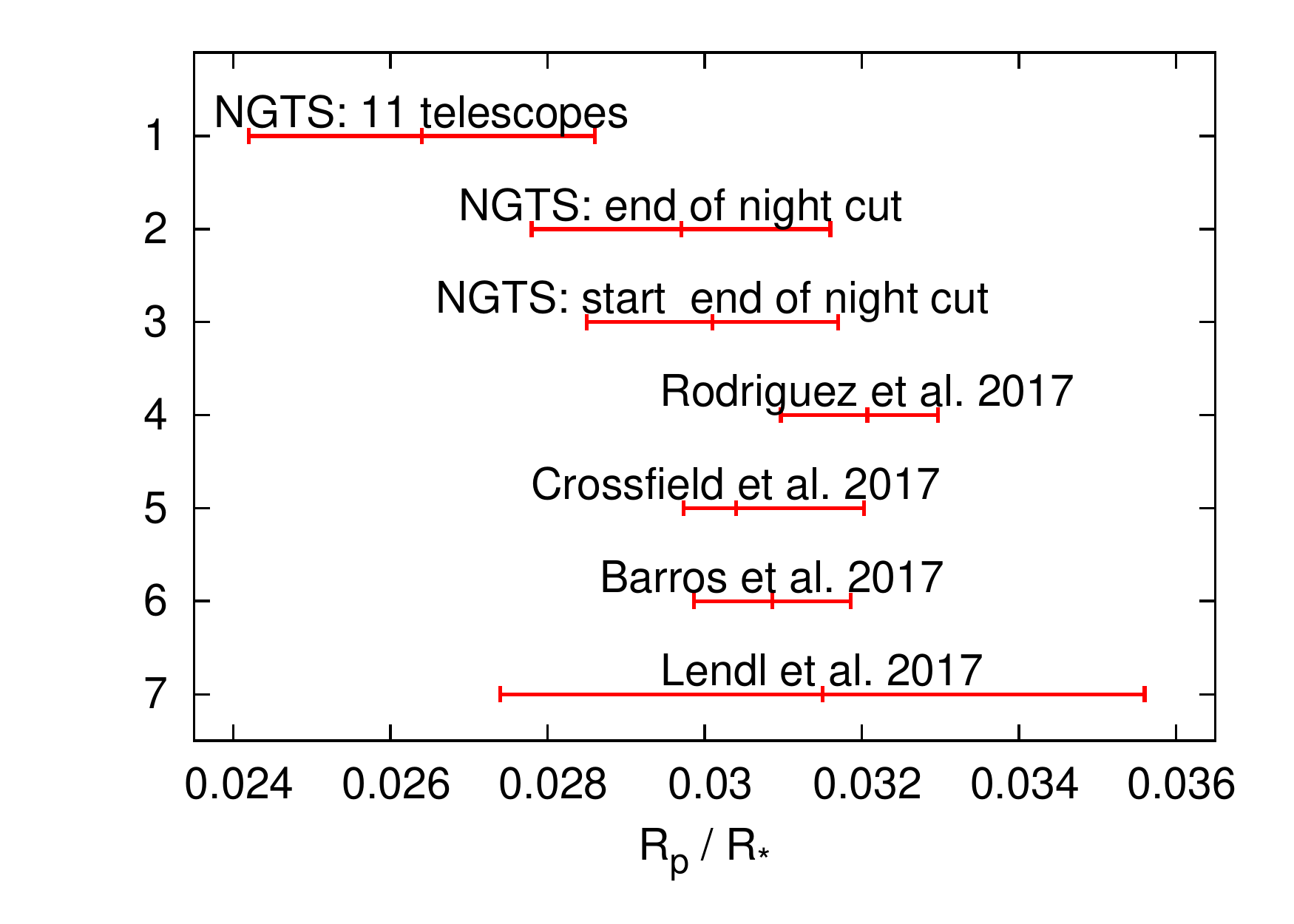} 
\caption{Comparison of our measured $R_p / R_*$ values with values in the published literature. The numbers on the y-axis correspond to those in Table~\ref{tab:radii}.
}
\label{fig:lit_radii} 
\end{figure}

In Table~\ref{tab:radii} and Fig.~\ref{fig:lit_radii} we compare the $R_p/R_*$ resulting from our fits to those previously published by others. Even without removing the high-airmass data, our result is in reasonable agreement with others (less than 2 $\sigma$, except for \cite{Rodriguez17}, with whose value ours is slightly more than 2 $\sigma$ discrepant). Excluding the high-airmass data at the end of the night from our fit results in an approx. 1 $\sigma$ change in the value of $R_p/R_*$. This new value is within about 1 $\sigma$ of all previously-published values.

The rms of the residuals to our eleven telescope fit is 850~ppm per minute or 240~ppm per half hour. These values fall to 777~ppm and 204~ppm respectively when removing the high-airmass data at the start of the night, and 657~ppm and 145~ppm when high-airmass data from both the beginning and the end are excluded.

\begin{table*}
\caption{Comparison of fitted planet-to-star radius ratio with previously-published values.}
\begin{center}
\begin{tabular}{lllcc}
\toprule
No. & Source   & $R_p/R_*$ & \multicolumn{2}{c}{Significance of difference w.r.t.}  \\ 
&&&full NGTS (1) & cut NGTS (2)\\
\midrule
(1)	&	NGTS (11 cams, full)            & $0.0264 \pm 0.0022$             & --            & 1.1 $\sigma$ \\
(2) &   NGTS (end of night cut)         & $0.0297 \pm 0.0019 $            & 1.1 $\sigma$  & -- \\
(3) &   NGTS (both ends of night cut)   & $0.0301 \pm 0.0016    $         & 1.4 $\sigma$  & 0.2 $\sigma$ \\
(4) &   \cite{Rodriguez17}    			& $0.0321^{+0.0009}_{-0.0011} $   & 2.6 $\sigma$  & 1.4 $\sigma$ \\
(5) &   \cite{Crossfield17}   		 	& $0.0304^{+0.0016}_{-0.0007} $   & 1.5 $\sigma$  & 0.4 $\sigma$ \\
(6) &   \cite{Barros17}   	 		    & $0.0309 \pm 0.0010 $            & 1.7 $\sigma$  & 0.6 $\sigma$ \\
(7) &   \cite{Lendl17}   	 		    & $0.0315 \pm 0.0041  $           & 1.0 $\sigma$  & 0.4 $\sigma$ \\
\bottomrule
\end{tabular}
\end{center}
\label{tab:radii}
\end{table*}

\section{Improving the ephemeris}

Since one of the motivations for these observations was to improve our knowledge of the planet's orbital ephemeris (Section~\ref{sec:intro}), we performed a series of fits designed to measure only the time of mid-transit. Our fitting procedure was similar to that described in Section~\ref{sec:combine}, but here we fixed the values of $a/R_*$, b, and $R_p/R_*$ to those determined from the \ktwo light curve \citep{Rodriguez17}. The epoch of mid-transit and a vertical offset (to account for imperfect flux normalisation) were the only parameters for which we fitted.

\begin{figure}
    \centering
    \includegraphics[width=\hsize]{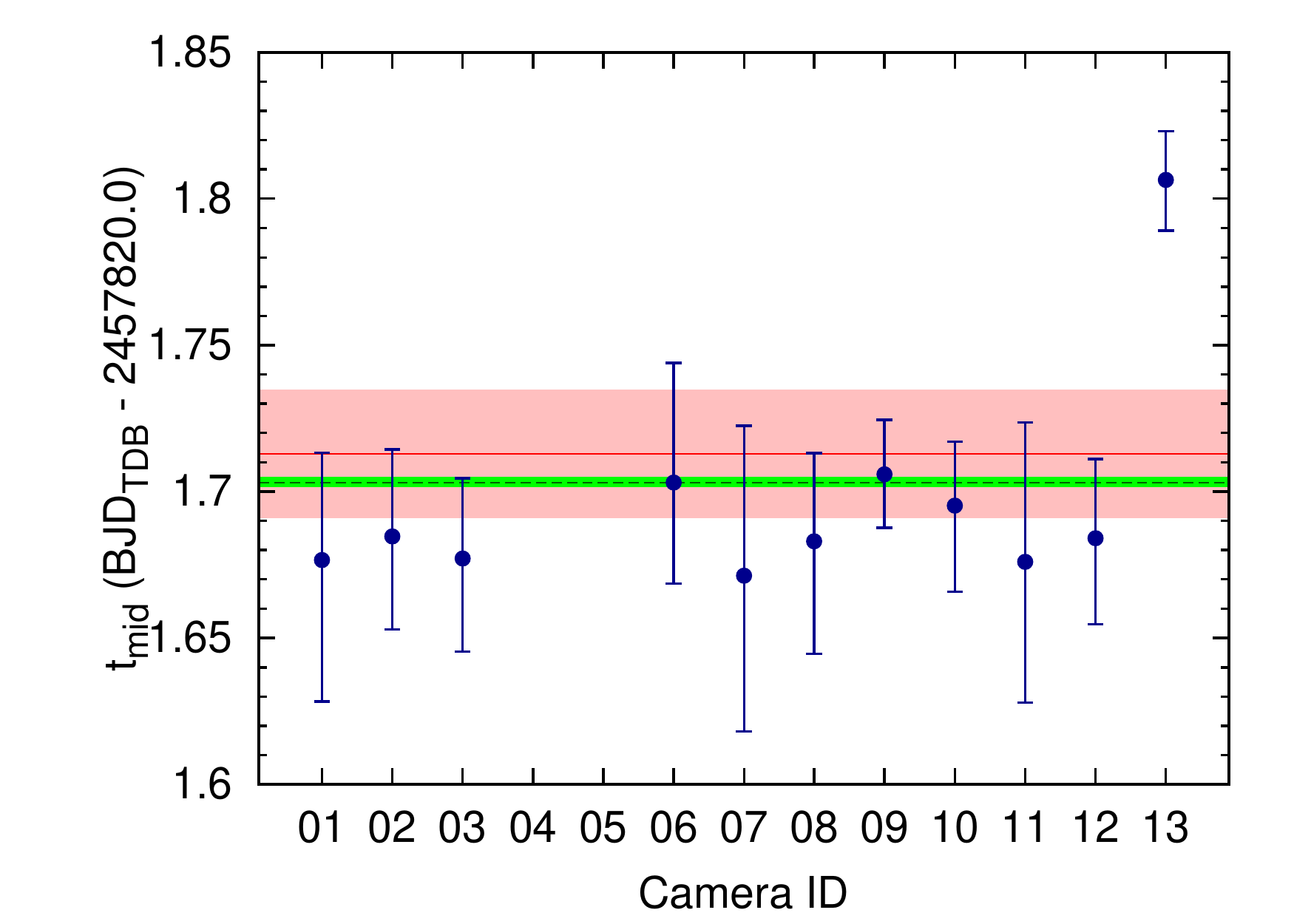}
    \caption{Fitted time of mid-transit for single-telescope light curves. The value and uncertainty of $t_\mathrm{mid}$ from the \ktwo ephemeris \citep{Rodriguez17} is indicated with a red solid line, and pink shaded region. The dashed green line, and region indicates the $t_\mathrm{mid}$ from the Eulercam observations \citep{Lendl17}.
    }
    \label{fig:epoch1}
\end{figure}

\begin{figure} 
\includegraphics[width=\hsize]{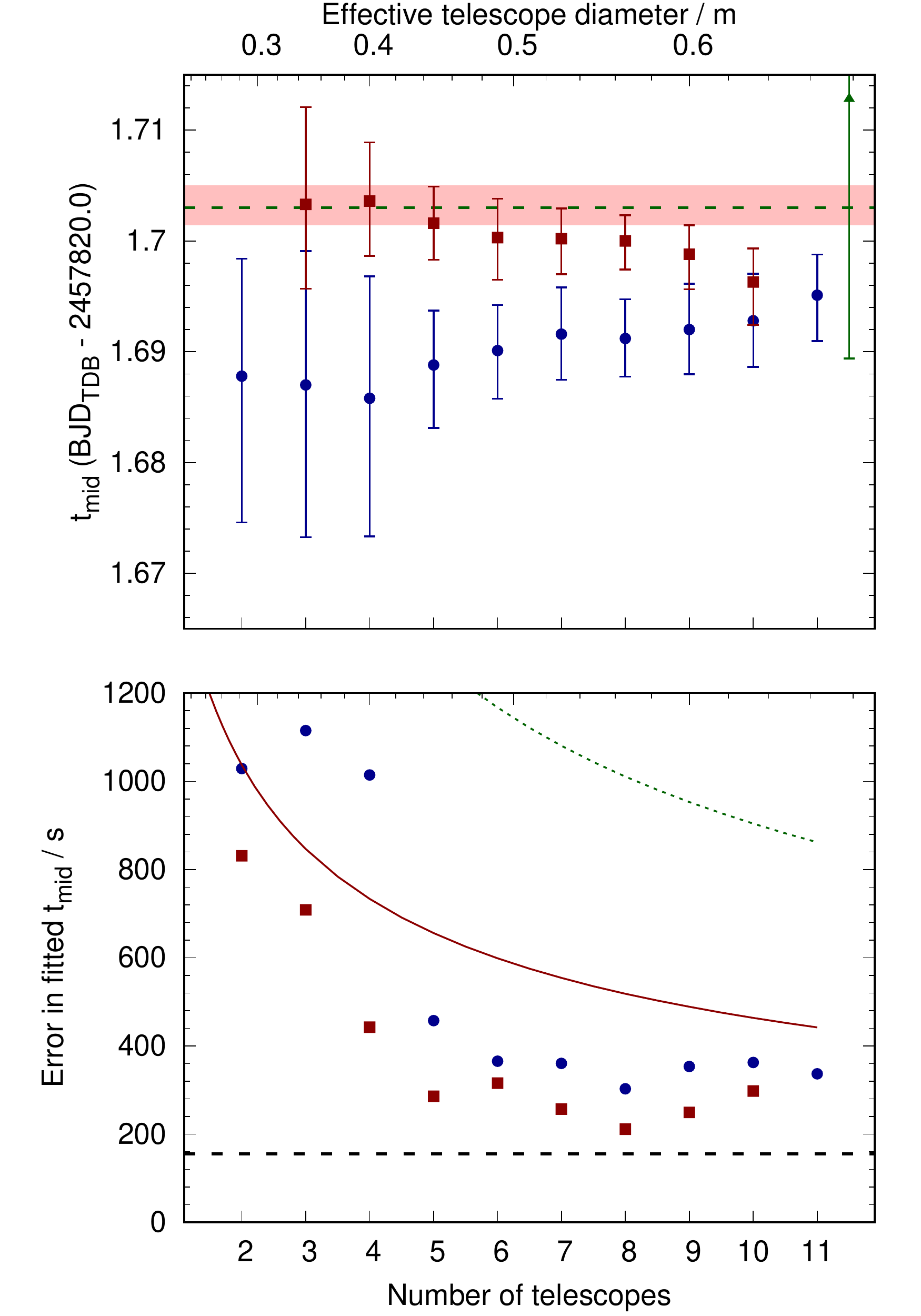}
\caption{Upper panel: fitted epoch of mid-transit as a function of number of telescopes combined. The blue circles represent the combinations starting with the two `best' (see text) light curves and adding successively worse light curves. The red squares represent the combinations starting with the two `worst' light curves, and adding successively better light curves. Note that the two-telescope point is a significant outlier, and is therefore not shown. The dashed green line and red shaded region indicate the best-fitting solution and 1 $\sigma$ uncertainties from observations with a 1-m telescope \citep{Lendl17}. The green triangle indicates the transit time and uncertainty predicted by the  ephemeris based on \ktwo photometry alone \citep{Rodriguez17}. Lower panel: uncertainty on transit epoch. The red squares and blue circles correspond to the same fits as the upper panel. As in Fig.~\ref{fig:combo}, the solid red line  represents the individual light curve with the smallest uncertainty, scaled by $1/\sqrt{n_\mathrm{tel}}$. The dotted green line is the same, but based on the mean uncertainty from an individual light curve. The dashed black line is the uncertainty obtained by observing the same transit using a 1-m telescope \citep{Lendl17}.
}
\label{fig:epoch2} 
\end{figure}

We performed a fit to each individual light curve, the results of which are shown in Fig.~\ref{fig:epoch1}. We also performed a series of fits, in which we incrementally added additional light curves. We used the same ranking of light curves as in Section~\ref{sec:radius}, and again performed two sets of fits, starting with both the `best' and the `worst' light curves. The results of these fits are shown in Fig.~\ref{fig:epoch2}.

Similarly to the radius-ratio case, we see that adding an increasing number of telescopes results in a better-determined transit time. The results show a greater departure from the simple white noise ($1/\sqrt{n_\mathrm{tel}}$ case than did the radius ratio. As the number of telescopes used increases, there is a relatively rapid improvement in our epoch determination until five telescopes, but then only modest improvement beyond that. Our value of the epoch from combining all eleven light curves is very close to that obtained with a 1-m telescope, both in the value and its precision \citep{Lendl17}. Our observations result in an epoch value that is significantly better determined than that from the \ktwo observations alone, reducing the 1$\sigma$ uncertainty from more than half an hour, to just 5.6 minutes. This demonstrates the power of such observations to improve the ephemeris, and hence the future observability, of transiting systems like \hd.

\section{Discussion and conclusions}
\label{sec:discuss}

\subsection{Summary and outlook for NGTS follow-up of TESS targets}

By observing the same target with multiple NGTS telescopes, and combining the resulting data, we are able to measure an exoplanet transit with a depth of just 1~mmag. The precision of this measurement rivals 1-m class telescopes such as Euler \citep{Lendl17}, even though the diameter of a single NGTS telescope is just 0.2~m. We are able to reduce the noise by binning data from multiple telescopes, as expected given that for bright stars the NGTS noise budget is dominated by scintillation \citep{NGTS}.

Our observations of \hd demonstrate the sensitivity of NGTS to shallow transits, particularly in `follow-up' rather than `survey' mode. TESS discovers a large number of transiting planets for which only one or two transits are observed with TESS \citep{cooke2018, Villanueva2019}. Observing additional transits with ground-based observations is crucial to refine system parameters, particularly the orbital ephemeris, and we have demonstrated here that NGTS is extremely-well suited to this task. 

For many bright targets, NGTS' wide field-of-view combined with its high photometric precision places it among the very best ground-based facilities for follow-up transit observations. This is because 1-m class telescopes, while perhaps offering similar photometric precision, typically have rather limited fields-of-view, resulting in few or no available reference stars of similar brightness to the target. Each NGTS telecope has a field-of-view of $2.8^{\circ} \times 2.8^{\circ}$, and thus plenty of reference stars for even bright targets.

Further observations in the multi-telescope mode employed here will allow us to build up experience of how photometric precision varies both with the number of telescopes used in the observations, and with target brightness. This will allow the selection of the optimal number of telescopes for a given target, improving the efficiency of telescope operations.

Since our observations of \hd, transits of several other targets have been successfully observed in multi-camera mode, with various numbers of cameras employed (\citealt{Lendl_TOI-222}; Jenkins et al., under review).

\subsection{Looking forward to PLATO}

Although seemingly very different types of transit survey, PLATO \citep{PLATO} and NGTS have several common characteristics which makes the analysis performed in this work relevant in the context of PLATO. PLATO is designed to detect the transits of Earth-sized planets in Earth-like orbits around Sun-like stars. However, such transits can only be detected by combining data from multiple PLATO telescopes. Datasets like the one analysed in this work, where the transit is shallow with respect to the noise level, therefore offer a platform to explore possible strategies for combining data from multiple telescopes in PLATO.

Both NGTS and PLATO consist of a number of identical individual telescopes, which are subject to sources of noise, some of which are common between multiple telescopes, and some of which act at the level of the individual telescopes (Table~\ref{tab:noise}). For instance, the PLATO telescopes share a common spacecraft platform and so jitter arising from the spacecraft pointing will affect all telescopes in a similar way. While the NGTS telescopes are mounted independently, they are all located in the same enclosure, and thus experience environmental and atmospheric effects in common.

PLATO will combine data from multiple telescopes taking non-simultaneous exposures (timing offsets are up to 18.75~s). Similarly, the NGTS exposures were not synchronised. The multi-telescope mode of NGTS offers the possibility of testing different approaches to combining / binning data from multiple cameras, with a view to optimising the performance of PLATO.

\begin{table*}
\caption{Comparison of noise sources in NGTS and PLATO photometry.}
\begin{center}
\begin{tabular}{lll} \toprule
   & NGTS & PLATO \\ \midrule
Individual telescope & pointing & TOU temperature\\
&focus& focus\\
&&position of target on CCD\\
\midrule
Common across all telescopes & airmass, temperature &  spacecraft pointing\\
&seeing& solar activity\\
&temperature&\\
\bottomrule
\end{tabular}
\end{center}
\label{tab:noise}
\end{table*}

\section*{Acknowledgements}

This work is based on data collected under the NGTS project at the ESO La Silla Paranal Observatory. The NGTS facility is operated by the consortium institutes with support from the UK Science and Technology Facilities Council (STFC) project ST/M001962/1. We thank our anonymous referee for their helpful comments, which helped to improve the manuscript.


\bibliography{refs2.bib, iau_journals.bib}

\appendix

\section{Fits and noise properties of individual NGTS light curves}
\label{appendix}

\begin{figure*}
    \centering
    \includegraphics[width=0.8\hsize]{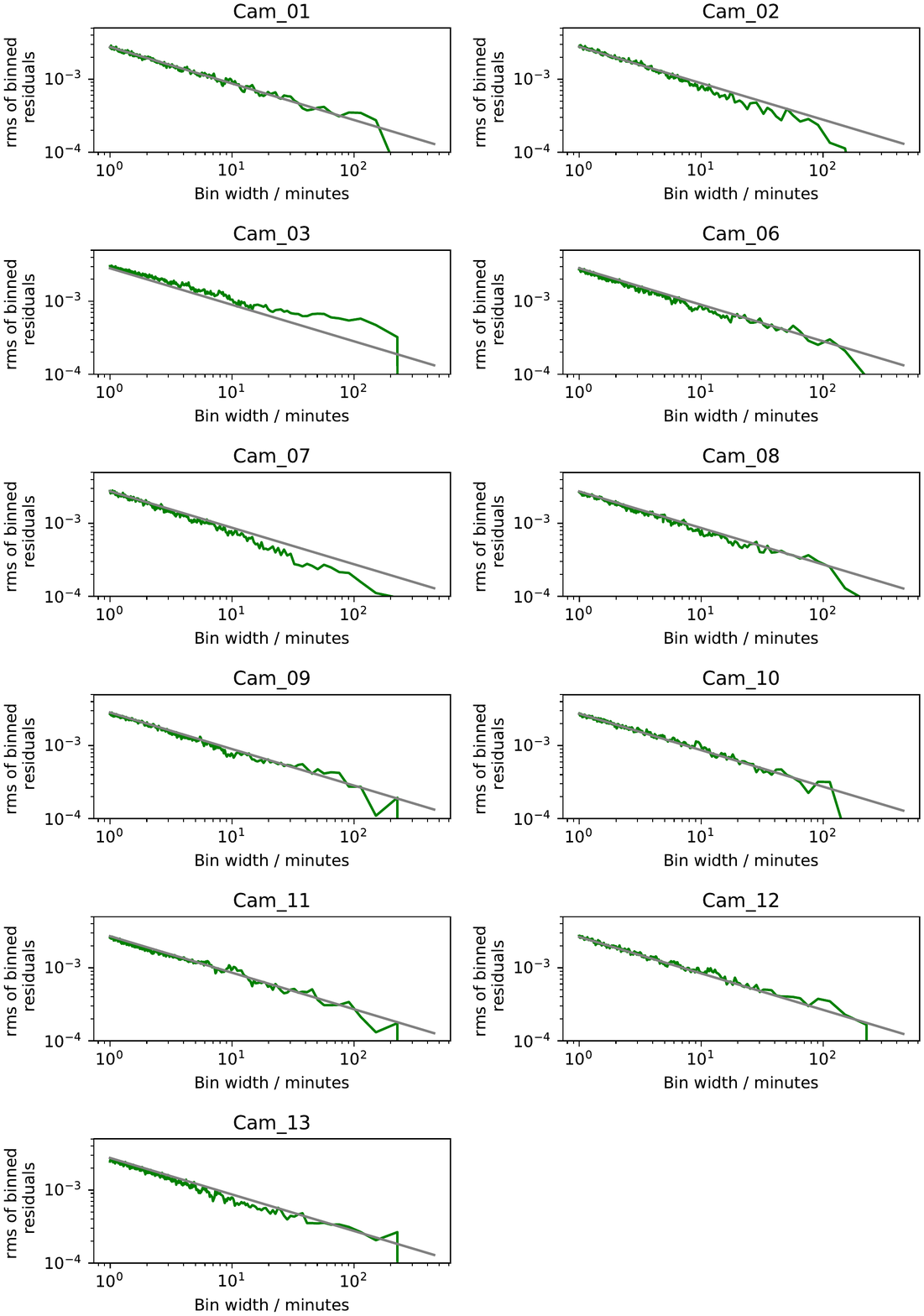}
    \caption{The rms of the binned residuals for each individual light curve (green curves). The plots here result from fits with fixed limb-darkening, and constrained impact parameter, but are virtually indistinguishable from those resulting from fits with the aforementioned parameters freely fitted. The white noise expectation, where the rms decreases in proportion to the square root of the bin size, is shown with a grey line in each panel.}
    \label{fig:rms}
\end{figure*}

As discussed in Section~\ref{sec:analysis_single}, we fitted each individual NGTS light curve separately. The residuals to these fits were analysed by binning them with a range of bin sizes, and determining the rms in each case. Fig.~\ref{fig:rms} shows the results of this analysis, and indicates that little-to-no residual systematic noise is present in the photometry, with the exception of camera 03.

\end{document}